# Synthesis and characterization of Ni$^0$ NPs immobilized on acid activated montmorillonite (Ni°-Mont) as an efficient and reusable heterogeneous catalyst for one-pot synthesis of biscoumarins under microwave irradiation

Behzad Zeynizadeh and Soleiman Rahmani*

Faculty of Chemistry, Urmia University, Urmia 5756151818, Iran
E-mail: s.rahmani@urmia.ac.ir

**Abstract**
The present paper describes in situ preparation of Ni° nanoparticles immobilized on acid activated montmorillonite K10 (Ni°-Mont) by impregnation of Ni(OAc)2 into nanopores of acid activated montmorillonite K10 followed by polyol reduction. Ni° NPs was highly dispersed into the initiated mesopores on the surface or inside swelled-interlayers of the examined clay with average size of 10-20 nm. Acid treatment of montmorillonite was carried out with HCl (4 M) under controlled conditions for generation of the desired pore sizes as a host of Ni° NPs. The prepared Ni°-Mont was used as a highly efficient heterogeneous catalyst for one-pot synthesis of biscoumarins in 85-95% yields. The reactions were carried out via Knoevenagel condensation of aromatic aldehydes with 4-hydroxycoumarin under microwave irradiation. The Ni°-Mont catalyst was reused at least for seven times without significant loss of its catalytic activity.

**Introduction**

In recent years, metal nanoparticles have attracted a great deal of attentions because of their distinct able properties like optical, electrical, antifungal, antibacterial, applications in the fields magnetic devices, sensors and drug delivery [1-5]. Metal NPs is emerging as very attractive catalysts when they possess a large surface to volume ratio compared to their bulk counterparts, due to quantum size and surface effects. This makes them particularly attractive for many organic reactions [6].

Ni$^0$ nanoparticles (Ni$^0$ NPs) due to nano-sized properties, applications in catalysis and magnetism as well as their chemical characteristics have gained the attention of chemists in many fields [7-9]. They are widely used in organic synthesis due to easy preparation, potent catalytic activity, using mild reaction conditions, high stability and easy recyclability in comparison to traditional Raney Ni. Recently, the use of nickel nanoparticles has been used in Suzuki coupling [10-12], reductive amination of aldehydes [13], hydrothermal Heck reaction [14-16], Hantzsch dihydropyridine synthesis [17], Knoevenagel condensation [18] and synthesis of pyran derivatives [19]. Therefore, preparation of metal NPs and to maintain them in smaller sizes is very important, because in most of cases they tend to agglomeration to form larger clusters. Agglomeration decreases the nanoparticle surface area for condensation and/or chemical reaction [20-22]. Using stabilizer/supports for synthesis of nanoparticles play an important role in preventing agglomeration, controlling nano size, shape as well as morphology [21, 23]. The most commonly used stabilizers are alkylammonium salts [24], organic ligand [25], polymer [26], carbon materials [27] and surfactants [28]. Some of mesoporous solid that are utilized as stabilizers of metal nanoparticles including of zeolites [29], resins [30], metal oxides [31] and clay minerals [32, 33]. Clay minerals of the smectite group like montmorillonite, K10 hectorite and saponite [32, 34, 35] are the suitable supports where metal nanoparticles can be stabilized into their interlamellar spaces. According to SEM image (Figure 2), The synthesized acid activated montmorillonite possesses high surface area and micro/meso pores which can uses for stabilizing various metal nanoparticles.

Biscoumarins (dicoumarols) are important and widely used precursors in synthesis of acridinediones (laser dyes) and heterocyclic compounds [36]. They have also received much attention in pharmaceutical studies. A literature review shows that many coumarins and dicoumarols exhibited a broad spectrum of biological activities such as antiinflammatory, antibacterial, antiviral, anticancer, anti-HIV, and antiproliferative properties [37-43]. It was also reported that some dicoumarol-based inhibitors showed an activity against gram-positive bacteria [44, 45]. Dicoumarols have been generally synthetized by refluxing 4-hydroxy-coumarin and various aldehydes in ethanol for several hours [46]. The literature review shows that they have also been synthetized in the presence of various catalysts such as molecular iodine [47], piperidine [48], DBU [49], n-dodecylbenzene sulfonic acid (DBSA) [50], LiClO$_4$ [51], sodium dodecyl sulfate (SDS) [52], Zn(proline)2 [53], tetra-n-butylammonium bromide (TBAB) [54], ion liquids [55], nano SiO$_2$Cl [56]. Although the titled protocols have their own advantages, however, most of them suffer from one or more disadvantages such as the prolonged reaction times, tedious workup procedures, unsatisfactory yields, high temperature and use of expensive catalysts. Thus the development of simple, efficient and ecofriendly methods using new catalysts for synthesis of biscoumarins is still demanded. In line with the outlined strategies, herein, we wish to introduce an efficient and novel method for synthesis of biscoumarins via one-pot multicomponent condensation of aromatic aldehydes with 4-hydroxycoumarin catalyzed by Ni$^0$ NPs immobilized on acid activated montmorillonite K10 as a green, robust and easily recoverable clay-based catalyst (Scheme 1).

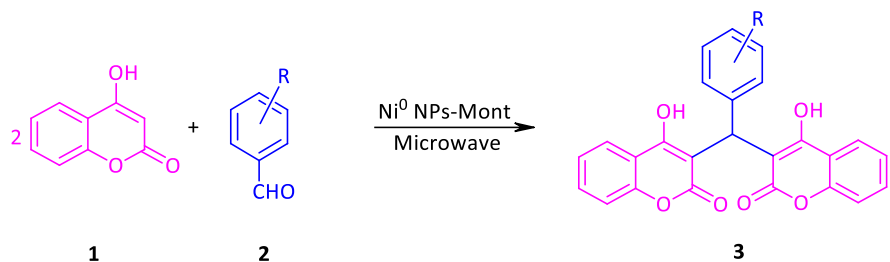

**Scheme 1**. Synthesis of biscoumarin (3) catalyzed by Ni$^0$ NPs-Mont system

**2-Experimental**

**2.1. Materials and methods**

All chemicals were purchased from Sigma-Aldrich and Merck chemical companies with the best quality and they were used without further purification. $^1$H/$^{13}$C NMR and FTIR spectra were recorded on 300 MHz Bruker Avance and Thermo Nicolet Nexus 670 spectrometers. Particle Size and morphology of materials were measured by using scanning electron microscopy (SEM) and energy dispersive X-ray spectroscopy (EDX). Melting points were recorded on Electrothermal 9100 melting point apparatus and uncorrected. All products are known and were characterized by comparison of their physical and spectral data with those of authentic samples. Yields refer to isolated pure products. TLC was applied for monitoring of the reactions over silica gel 60 F254 aluminum sheet. Microwave irradiation was carried out in a domestic microwave oven LG MS-2344B (1250 W). Montmorillonite K10 was purchased from Sigma-Aldrich company art No. 69866 (pH ~ 3-4, surface area: 250 m$^2$/g).

**2.2 Preparation of homoionic Na$^+$-exchanged montmorillonite (Na$^+$-Mont)**

In a beaker (250 mL) containing distilled water (200 mL), montmorillonite K10 (5 g) was added and the mixture was stirred vigorously at room temperature for 20 h. The mixture was then allowed to settle and the aqueous phase was decanted. To the obtained solid residue, an aqueous solution of NaCl (2 M, 200 mL) was added and the mixture was continued to stirring for 2 h at room temperature. The aqueous phase was decanted and the solid residue was again charged with an aqueous solution of NaCl (2 M, 200 mL). After stirring for 2 h at room temperature, the aqueous phase was decanted. The procedure was repeated for additional 2 times. Finally, the solid residue was washed frequently with distilled water until the conductivity of the liquid filtrate reaches to the conductivity of distilled water. The solid residue was dried at 50 °C under air atmosphere to afford homoionic Na$^+$-exchanged montmorillonite [33].

**2.3. Preparation of acid-activated montmorillonite [H$^+$-Mont]**

To a round-bottom flask (250 mL) containing an aqueous solution of HCl (4 M, 100 mL), Na$^+$-Mont (5 g) was added. The mixture was stirred under reflux conditions for 2 h. After cooling, the aqueous phase was decanted and the residue was washed frequently with distilled water. When the liquid filtrate was free from Cl⁻ (test with $AgNO_3$), the solid material on filter paper was collected and dried at 50 °C to afford acid-activated montmorillonite.

### 2.4. Preparation of $Ni^0$ NPs immobilized on acid-activated montmorillonite [$Ni^0$-Mont]

In a beaker (100 mL) containing an aqueous solution of $Ni(OAc)_2 \cdot 2H_2O$ (0.05 M, 10 mL), acid-activated montmorillonite (0.5 g) was added slowly and the resulting mixture was stirred vigorously at room temperature for 6 h. The aqueous phase was then evaporated under reduced pressure. The dried Mont-$Ni(OAc)_2$ was dispersed in ethylene glycol (50 mL) within a double necked round-bottom flask. The mixture was stirred at reflux for 6 h under nitrogen atmosphere. Ethylene glycol was then removed by decantation and the solid residue was washed with MeOH up to complete removing of ethylene glycol. The solid material was dried through nitrogen flowing (12 h) at 40 °C to afford $Ni^0$ NPs immobilized on acid activated montmorillonite.

### 2.5. A general procedure for synthesis of biscoumarins using $Ni^0$ NPs immobilized on acid activated montmorillonite

A mixture of 4-hydroxycoumarin (2 mmol), aromatic aldehyde (1 mmol) and $Ni^0$-Mont (0.02 g) were mixed together and irradiated in a domestic microwave oven (850 W) for 5 min. After completion of the reaction (monitored by TLC), the mixture was cooled to the room temperature. EtOAc (5 mL) was then added and the mixture was stirred for 5 min. The catalyst was removed from the reaction mixture by a filter paper. The filtrate was evaporated and the solid material was recrystallized from hot ethanol to obtain the pure biscoumarins as white solids (Table 2, 85-95% yields).

**3,3'-(phenylmethylene)-bis-(4-hydroxy-2*H*-chromen-2-one) (Table 2, entry 1)**

FTIR (KBr, υ Cm⁻¹) 3400, 3067, 2737, 1659, 1609, 1564, 1336, 1089, 755; ¹H NMR (300 MHz, $CDCl_3$) δ (ppm) 6.11 (s, 1H, CH), 7.25-8.07 (m, 12H, ArH), 11.32 (s, 1H, OH), 11.55 (s,1H, OH). ¹³CNMR (75 MHz, $CDCl_3$) δ 36.16, 103.88, 105.62, 116.42, 116.65, 124.39, 124.91, 126.89, 128.77, 132.88, 135.17, 152.27, 164.60, 165.02.

**3,3'-(2,4-dichlorophenylmethylene)bis(4-hydroxy-2*H*-chromen-2-one) (Table 2, entry 2)**

FTIR (KBr, υ Cm⁻¹) 3068, 2719, 2594, 1649, 1565, 1497, 1097, 759; ¹H NMR (300 MHz, $CDCl_3$) δ (ppm) 6.12 (s, 1H, CH), 7.24-8.02 (m, 11H, ArH), 11.39 (s, 2H, OH); ¹³C NMR (75 MHz, $CDCl_3$) δ 35.46, 116.51, 124.48, 124.87, 126.9, 130.33, 132.92, 134.1, 152, 164.22, 166.8.

**3,3'-(4-Chlorophenylmethylene)-bis-(4-hydroxy-2*H*-chromen-2-one) (Table 2, entry3)**

FTIR (KBr, υ Cm⁻¹) 3400, 3073, 2730, 1664, 1614, 1565, 1494, 1344, 1094, 770; ¹H NMR (300 MHz, $CDCl_3$): δ (ppm) 6.04 (s, 1H, CH), 7.15-8.06 (m, 12H, ArH), 11.32 (s, 1H, OH), 11.55 (s,1H, OH). ¹³C NMR (75 MHz, $CDCl_3$) δ 35.7, 103.69, 105.24, 116.34, 124.41, 125, 127.97, 128.77, 132.77, 133.04, 133.83, 152.26, 164.53, 166.02.

**3,3'-(2-chlorophenylmethylene)bis(4-hydroxy-2*H*-chromen-2-one) (Table 2, entry 4)**

FTIR (KBr, υ Cm⁻¹) 3069, 2720, 1649, 1556, 1498, 1444, 1340, 1097, 760; ¹H NMR (300 MHz, $CDCl_3$):δ (ppm) δ 6.01 (s, 1H, CH), 7.27-8.04 (m, 12H, ArH), 10.93 (s, 1H, OH), 11.63 (s, 1H, OH). ¹³C NMR (75 MHz, $CDCl_3$) δ 35.72, 116.59, 124.42, 124.87, 126.76, 128.59, 129.24, 130.81, 132.83, 133.49, 162.33.

**3,3'-(4-bromophenylmethylene)bis(4-hydroxy-2*H*-chromen-2-one) (Table 2, entry 5)**

FTIR (KBr, υ Cm⁻¹) 3428, 3068, 2930, 1663, 1612, 1563, 1490, 1445, 1342, 1315, 1200, 1096, 1022, 767; ¹H NMR (300 MHz, $CDCl_3$) δ (ppm) 6.02 (s, 1H, CH), 7.09-8.06 (m, 12H, ArH), 11.31 (s, 1H, OH), 11.53 (s, 1H, OH); ¹³C NMR (75 MHz, $CDCl_3$) δ 35.86, 103.64, 105.18, 116.36, 116.66, 120.87, 124.4, 124.97, 128.3, 131.69, 134.4, 152.28, 162.34, 164.63, 166.02.

**3,3'-(4-Methoxybenzylidene)bis(4-hydroxy-2*H*-chromen-2-one) (Table 2, entry 6)**

FTIR (KBr, υ Cm⁻¹) 3064, 2944, 2842, 2727, 1666, 1612, 1562, 1507, 1447, 1343, 1309, 1094, 768; ¹H NMR (300 MHz, $CDCl_3$) δ (ppm) 3.80 (s, 3H, OMe), 6.05 (s, 1H, CH), 6.84-8.05 (m, 12H, ArH), 11.34 (s, 1H, OH), 11.52 (s, 1H, OH); ¹³C NMR (75 MHz, $CDCl_3$) δ 33.37, 76.62, 77.04, 77.47, 110.93, 116.5, 116.7, 120.4, 123.5, 124.3, 124.7, 128.2, 128.4, 132.44, 152.13, 157.56, 163.7, 165.2.

**3,3'-(2-methoxyphenylmethylene)bis(4-hydroxy-2*H*-chromen-2-one) (Table 2, entry 7)**

FTIR (KBr, υ Cm$^{-1}$) 3412, 3069, 2981, 2724, 1654, 1620, 1550, 1494, 1448, 1337, 1305, 1094, 758; $^1$H NMR (300 MHz, CDCl$_3$) δ (ppm) 3.5 (s, 3H, OMe), 6.09 (s, 1H, CH), 6.98-8.04 ( m, 12H, ArH), 11.22 (2s, 2H, OH); $^{13}$C NMR (75 MHz, CDCl$_3$) δ 35.48, 76.64, 77.06, 77.49, 104.1, 105.7, 113.9, 116.61, 116.89, 124.35, 124.86, 126.9, 127.6, 152.25, 152.47, 164.51, 165.69, 169.25.

**3,3'-(p-tolylmethylene)bis(4-hydroxy-2*H*-chromen-2-one) (Table 2, entry 8)**

FTIR (KBr, υ Cm$^{-1}$) 3100, 2992, 2919, 1665, 1614, 1563, 1502, 1440, 1094, 765; $^1$H NMR (300 MHz, CDCl$_3$) δ (ppm) 2.34 (s, 3H, CH$_3$), 6.07 (s, 1H, CH), 7.12-8.04 (m, 12H, ArH), 11.34 (s, 1H, OH), 11.48 (s, 1H, OH) $^{13}$C NMR (75 MHz, CDCl$_3$) δ 20.97, 35.87, 116.61, 116.93, 124.37, 124.84, 126.36, 129.33, 132.77, 136.46, 152.47, 162.34, 164.5, 165.67.

**3,3'-(4-Hydroxyphenylmethylene)bis(4-hydroxy-2*H*-chromen-2-one) (Table 2, entry 9)**

FTIR (KBr, υ Cm$^{-1}$) 3361, 3072, 2731, 1660, 1614, 1565, 1509, 1443, 1096, 762; 1H NMR (300 MHz, CDCl3) δ (ppm) 6.02 (s, 1H, CH), 6.77-8.02 ( m, 12H, ArH), 11.48 (brs, 2H, OH); $^{13}$C NMR (75 MHz, CDCl$_3$) δ (ppm) 35.49, 104.16, 115.64, 116.61, 124.37, 126.78, 127.75, 152.41, 154.66, 162.32.

**3,3'-(4-nitrophenylmethylene)bis(4-hydroxy-2*H*-chromen-2-one) (Table 2, entry 10)**

FTIR (KBr, υ Cm$^{-1}$) 3427, 3075, 2935, 1652, 1562, 1343, 1095, 762; $^1$H NMR (300 MHz, CDCl$_3$) δ (ppm) 6.14 (s, 1H, CH), 7.21-8.17 (m, 12H, ArH), 11.39 (s, 1H, OH), 11.58 (s, 1H, OH); $^{13}$C NMR (75 MHz, CDCl$_3$) δ 36.18, 103.22, 104.62, 116.26, 116.73, 121.76, 122.12, 124.5, 125.17, 129.59, 132.7, 133.33, 137.9, 148.75, 152.35, 162.34, 166.59, 166.9.

**3,3'-(2-nitrophenyl)methylene)bis(4-hydroxy-2*H*-chromen-2-one) (Table 2, entry 11)**

FTIR (KBr, υ Cm$^{-1}$) 3433, 3076, 2722, 1652, 1618, 1561, 1532, 1450, 1352, 1309, 1099, 761; $^1$H NMR (300 MHz, CDCl$_3$) δ (ppm) 6.64 (s, 1H, CH), 7.27-8.01 (m, 12H, ArH), 11.54 (s, 2H, OH); $^{13}$C NMR (75 MHz, CDCl$_3$) δ 33.86, 103.71, 116.66, 124.57, 125.03, 128.19, 129.49, 131.9, 133.2, 149.76.

**3,3́-(N-pyridylmethylene)bis-(4-hydroxy-2*H*-chromen-2-one) (Table 2, entry 12)**

FTIR (KBr, υ Cm$^{-1}$) 3447, 3075, 2880, 1663, 1618, 1543, 1500, 1407, 1102, 757; $^1$H NMR (DMSO-d$_6$) δ (ppm) 6.47 (s, 1H, CH), 7.21-8.68 (m, 12H, ArH); $^{13}$C NMR (75 MHz, DMSO-d$_6$) δ (ppm) 38.26, 101.88, 116.20, 119.82, 123.65, 124.67, 125.67, 132.04, 141.35, 153.15, 164.53, 165.34, 168.59.

## 3. Results and Discussion

### 3.1 Synthesis and characterizations of Ni$^0$ NPs immobilized on acid-activated montmorillonite K-10

Aligned to the outlined strategies and due to lack of information for montmorillonite-catalyzed synthesis of biscoumarints, we therefore prompted to investigate the titled transformation by a novel Ni$^0$ NPs immobilized on acid activated montmorillonite K10 as a reusable metal nanoparticle clay based catalyst. The study was started primarily by synthesis of the clay catalyst in a five steps procedure: i) preparation of swelled (hydrated) montmorillonite through the vigorous stirring of commercially available montmorillonite K10 in distilled water. Similar to many other clay minerals, montmorillonite K10 absorb water between their interlayer space, which move apart and the clay swells. This swelling is caused by the hydration of interlayer cations which not only contributes to the acidity of montmorillonite but also it considerably expand the montmorillonite due to water penetrating the interlayer spaces, [6] ii) preparation of homoionic Na+-exchanged montmorillonite by adequate stirring of the swelled montmorillonite in an aqueous solution of NaCl. These sodium cations in the interlayer are highly exchangeable, which making montmorillonite able to accommodate various guest molecules (metal complexes, metal NPs and protons) in its interlayer space, [6] iii) preparation of acid-activated montmorillonite by stirring of homoionic Na$^+$-exchanged montmorillonite in an aqueous solution of HCl, iv) preparation of the immobilized Ni(OAc)$_2$ on acid-activated montmorillonite by a simply mixing and stirring of acid-activated montmorillonite in an aqueous solution of nickel acetate, and finally v) reduction of nickel acetate to Ni$^0$ NPs with ethylene glycol to obtain the final montmorillonite-based Ni$^0$ NPs (Scheme 2).

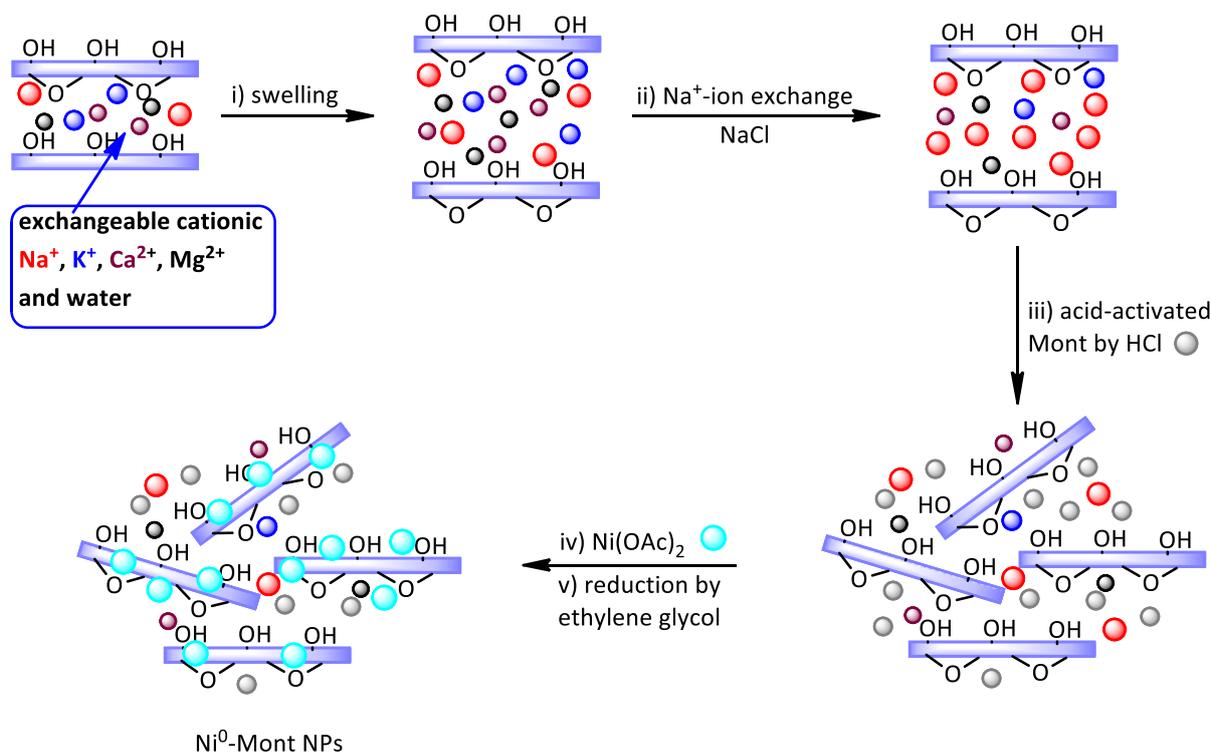

**Scheme 2**. Preparation of Ni$^0$ NPs-Mont system

After synthesis of Ni$^0$ NPs-Mont composite, the size and morphology of Mont k10, acid activated Mont and Ni$^0$-Mont was studied using FTIR, scanning electron microscopy and energy dispersive X-ray spectroscopy (EDX).

The illustrated FTIR spectra (Figure 1) exhibited that Mont K-10 shows a strong absorption band at 1054 cm$^{-1}$ indicating Si-O stretching vibrations of tetrahedral sites. The bands at 522 and 464 cm$^{-1}$ are also attributed to Si–O–Al and Si–O–Si bending vibrations. Through the acid activation of montmorillonite, the band at 1054 cm$^{-1}$ is shifted to 1080 cm$^{-1}$. A small band near 795 cm$^{-1}$ is due to amorphous silica. Mont K10 also shows absorption bands at 3620 and 1635 cm$^{-1}$ that are attributed to stretching and bending vibrations of OH groups of Al–OH. By immobilization of Ni$^0$ NPs, some changes in absorption bands of the parent spectra are observable.

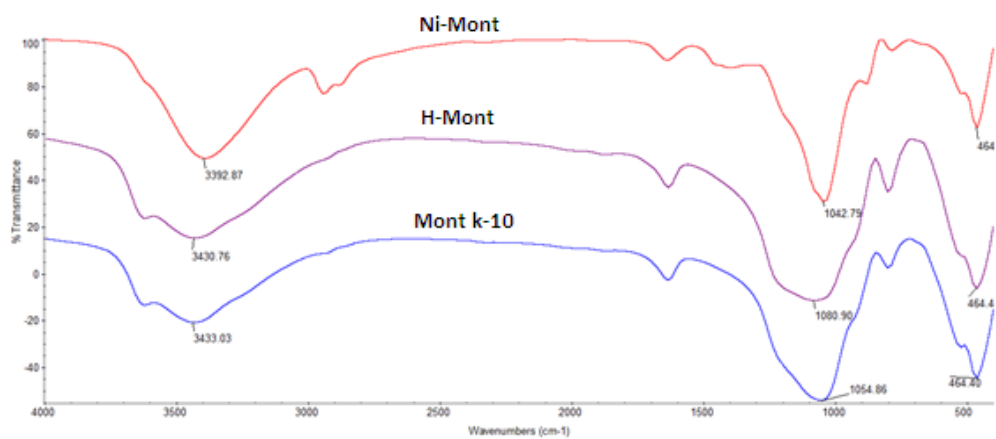

**Figure 1**. FTIR spectra of Mont K-10, acid-activated Mont and Ni$^0$ NPs-Mont

SEM images and EDX spectra of Mont k-10 and acid-activated Mont are shown in Figure 2. Comparison of images (**a** and **c**) shows that via the acid activation, the conjunct sheets of Mont k10 are exfoliated to tiny segments. This transformation is interpreted by elimination of major content of Al from octahedral sites of the aggregated sheets through the reaction with HCl. In addition, by the elimination of Al contents, new pores on the surface and internal sheets of montmorillonite are produced leading to raise surface area.

SEM images of the immobilized Ni NPs on acid-activated Mont (Figure 3) reveal that the immobilized particles have spherical shapes and well dispersed in the pores and on surface of tiny segments with a particle size of 10-20 nm. EDX analysis also indicates that Ni$^0$ NPs as well as other elements are present in the final catalyst.

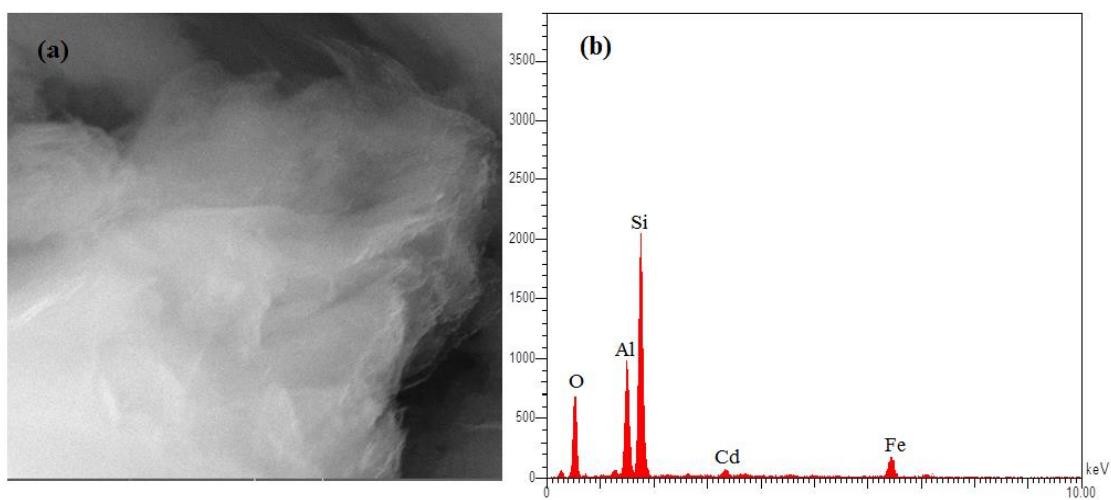

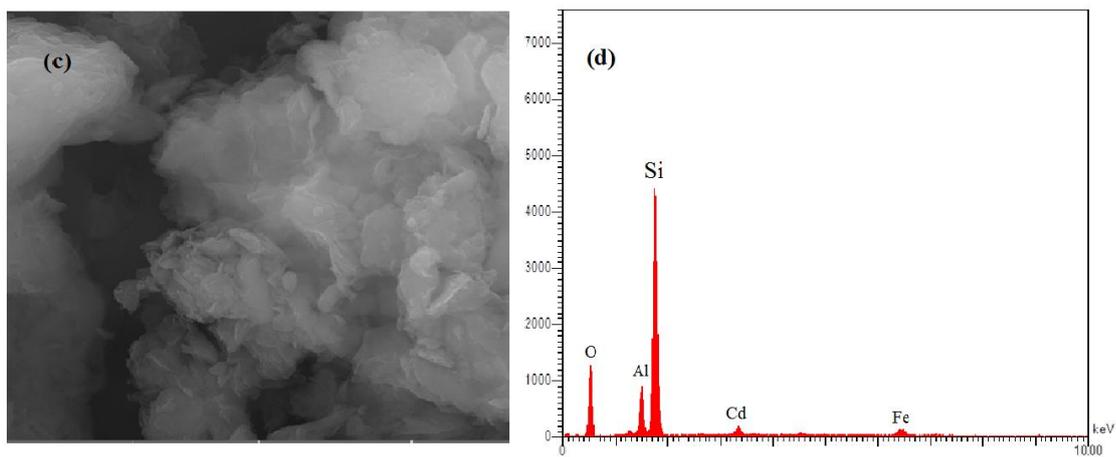

**Figure 2**. a) SEM image of Mont K-10, b) EDX spectra of Mont K-10, c) SEM image of acid-activated Mont, and d) EDX spectra of acid-activated Mont

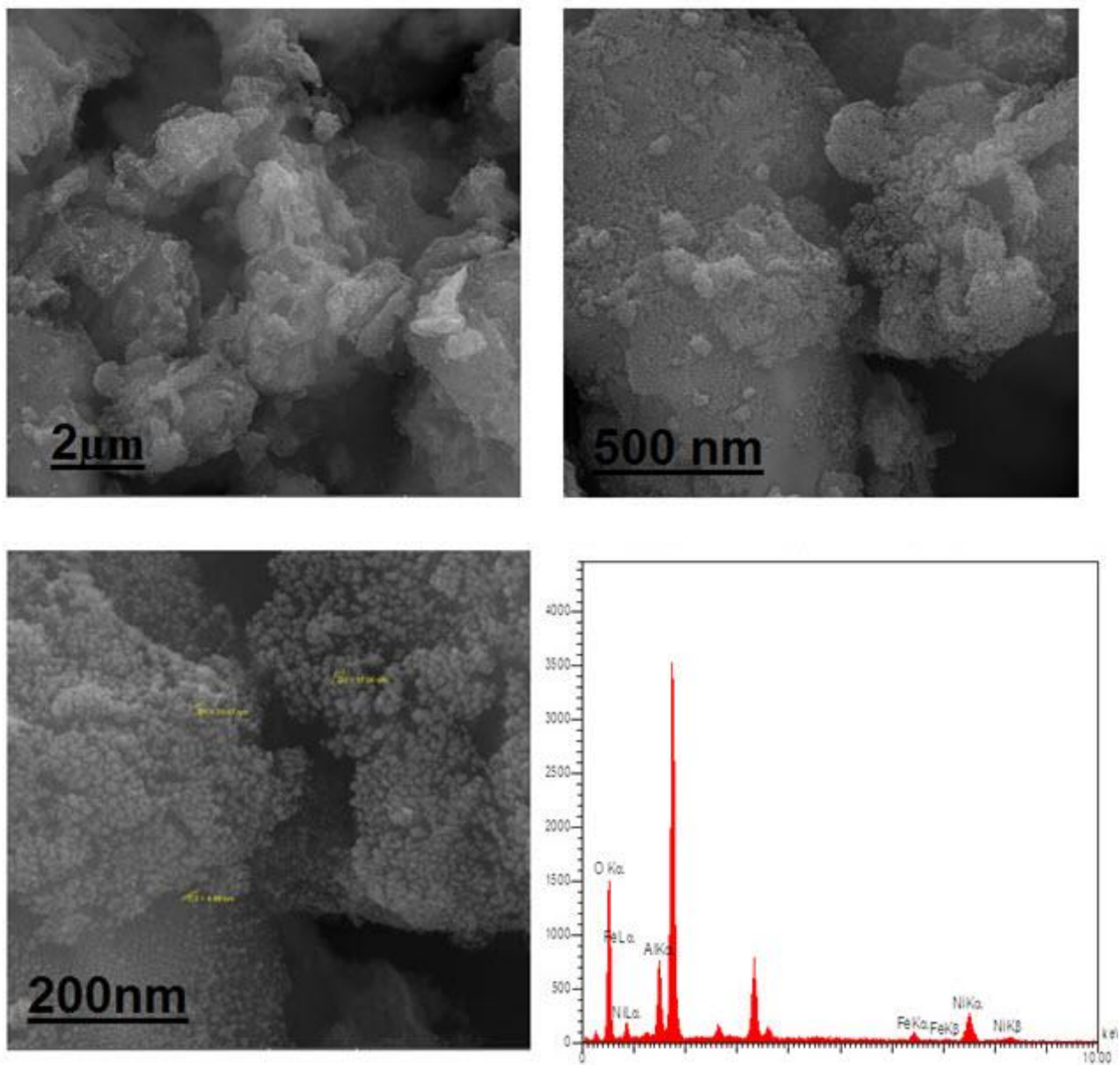

**Figure 3**. SEM and EDX spectra of Ni0 NPs on acid-activated Mont

### 3.2. Synthsis of biscoumarins catalyzed by Ni$^0$ NPs-Mont composite

At the next, the catalytic activity of Ni$^0$ NPs-Mont composite was studied by synthesis of biscoumarins via Knoevenagel condensation of aromatic aldehydes and 4-hydroxycoumarin. In order to optimize reaction conditions, progress of the reaction of 4-hydroxycoumarin (2 mmol) and 4-chlorobenzaldehyde (1 mmol) in the presence of Ni$^0$-Mont was investigated under different conditions. The results of this investigation were illustrated in Table 1. As seen, performing of the model reaction inside various solvents under reflux conditions led to poor yield of the final product. However, irradiation under microwaves (810 W) at solvent-free conditions, dramatically accelerated the rate of reaction to afford 3,3'-(4-chlorophenylmethylene)-bis-(4-hydroxy-2*H*-chromen-2-one) in 95% yield within 5 min. It is notable that using high molar equivalents of Ni$^0$-Mont did not accelerate the rate of reaction. In addition, completion of the model reaction in the presence of lower amounts of Ni$^0$-Mont composite was prolonged the reaction time. So, the conditions mentioned in entry 5 (Table 1) were selected as the optimum for complete conversion of 4-hydroxy-coumarin (2 mmol) and 4-chlorobenzaldehyde to the corresponding biscoumarin.

**Table 1**. Optimization experiments for synthesis of 3,3'-(4-chlorophenylmethylene)-bis-(4-hydroxy-2H-chromen-2-one) in the presence of Ni0-Mont composite

| Entry | Ni$^0$-Mont (mg) | Solvent (2 mL) | Condition | Time (min) | Yield (%) |
|---|---|---|---|---|---|
| 1 | 20 | EtOH | Reflux | 40 | 35 |
| 2 | 20 | H2O | Reflux | 120 | 30 |
| 3 | 20 | EtOH-H2O (1:1) | Reflux | 30 | 70 |
| 4 | 20 | CH3CN | Reflux | 120 | 20 |
| 5 | 20 | Solvent-free | Microwave | 5 | 95 |
| 6 | 30 | Solvent-free | Microwave | 5 | 95 |
| 7 | 10 | Solvent-free | Microwave | 10 | 83 |

The scope and generality of this synthetic protocol was more studied by synthesis of structurally diverse biscoumarins through the reaction of 4-hydroxycoumarin and substituted aromatic aldehydes under solvent-free and microwave irradiation. The illustrated results in Table 2 show the general trends and versatility of this synthetic protocol. As shown, all reactions were carried out within 5-15 min to afford the products in high to excellent yields.

**Table 2**. One-pot synthesis of biscoumarins catalyzed by Ni$^0$ NPs-Mont composite[a]

| Entry | Ar | Ni0-Mont (mg) | Time (min) | Yield (%)[b] | m.p. found | m.p. reported |
|---|---|---|---|---|---|---|
| 1 | C6H4 | 20 | 5 | 95 | 213-217 | 215[57] |
| 2 | 2,4-ClC6H4 | 20 | 10 | 90 | 265 | – |
| 3 | 4-ClC6H4 | 20 | 5 | 95 | 255-257 | 258-259[47] |
| 4 | 2-ClC6H4 | 20 | 10 | 85 | 202-204 | 198-199 |
| 5 | 4-BrC6H4 | 20 | 5 | 90 | 264-267 | 264-266[47] |
| 6 | 4-MeOC6H4 | 20 | 5 | 90 | 247-249 | 249-250[47] |
| 7 | 2-MeOC6H3 | 20 | 5 | 82 | 226-228 | – |
| 8 | 4-MeC6H4 | 20 | 5 | 90 | 267-269 | 269-270[58] |
| 9 | 4-OHC6H4 | 20 | 5 | 85 | 222-225 | 222-224 |
| 10 | 4-O2NC6H4 | 20 | 5 | 90 | 232-234 | 238-239[47] |
| 11 | 2-O2NC6H4 | 20 | 15 | 90 | 200-202 | 200-202[57] |
| 12 | 4-Me2NC6H4 | 20 | 5 | 90 | 214-215 | 214-215 |

a All reactions were carried out under microwave irradiation (850 W). b Yields refer to isolated pure products.

Although the exact mechanism of this synthetic protocol is not clear, however, we think that a depicted mechanism (Scheme 2) maybe play a role in the formation of biscoumarins. The mechanism shows that through the polarization of formyl group with Ni⁰-Mont composite, the formation of intermediate (**I**) was facilitated. Subsequently, by the reaction of 4-hydroxycoumarin with Intermediate (**I**), the intermediate (**II**) was produced. Finally, Michael addition of intermediate (**II**) with second molecule of 4-hydroxycoumarin affords the final biscoumarin.

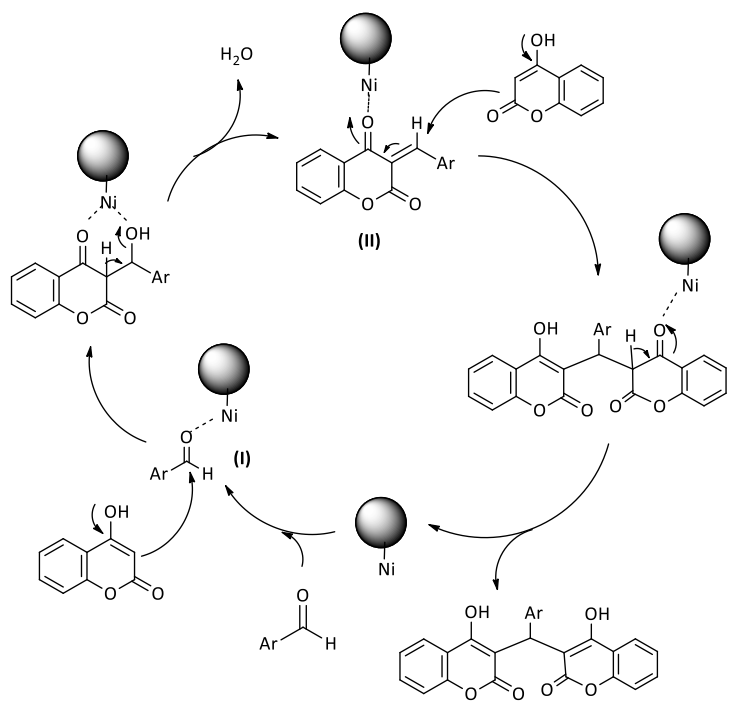

**Scheme 3**. Proposed mechanism for synthesis of biscoumarins

### 3.3. Recycling of Ni⁰ NPs-Mont Catalyst

In practical purposes, recyclability of the catalyst is highly desirable. To investigate this issue, reusability of Ni⁰ NPs-Mont composite was examined in Knoevenagel condensation of the model reaction at solvent-free under microwave irradiation. After completion of the reaction, the catalyst was separated by a filter paper and washed with MeOH and then dried under air atmosphere. The vessel of reaction was again charged with 4-chlorobenzaldehyde, 4-hydroxycoumarin and the recycled Ni⁰ NPs-Mont composite to run the condensation reaction for a second time. The examinations showed that the catalyst can be reused for 7 times without significant loss of its catalytic activity (Figure 4).

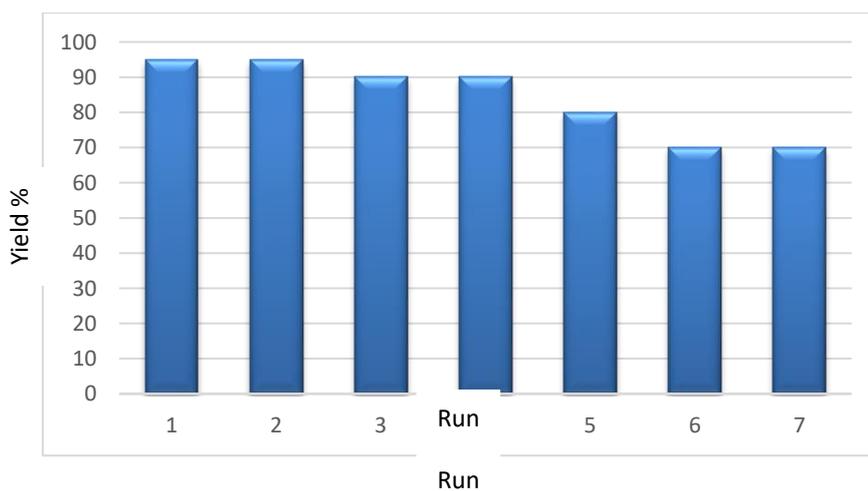

**Figure 4**. Reusability of Ni[0] NPs-Mont composite for synthesis of biscoumarins

**Conclusions**

In this paper, we have represents an easy protocol for immobilization of Ni[0] NPs into the initiated mesopores on the surface or spaces between interlayers of montmorillonite K10 (Ni[0]-Mont composite) as a new and reusable caly-based generation for the efficient one-pot Knoevenagel condensation of 4-hydroxycoumarin with structurally diverse aromatic aldehydes under solvent-free and microwave irradiation. Short reaction times, high to excellent yield of the products, reusability of Ni[0] NPs-Mont catalyst for 7 catalytic cycles as well as the benefits of solvent-free conditions are the advantages that make this protocol a prominent choice for synthesis of biscoumarins.